\documentclass[journal=jacsat,manuscript=article]{achemso}
\usepackage{subcaption}
\usepackage{graphicx}
\usepackage{array}
\usepackage{adjustbox}
\usepackage{longtable}
\usepackage{xcolor}
\usepackage[normalem]{ulem} 
\usepackage[version=3]{mhchem} 

\author{Juan Gomez Quispe}
\affiliation[Universidade Federal do ABC]
{Center of Natural and Human Sciences, Federal University of ABC, Santo Andre, Sao Paulo, Brazil}

\author{Matheus Medina}
\affiliation[Universidade Federal do ABC]
{Center of Natural and Human Sciences, Federal University of ABC, Santo Andre, Sao Paulo, Brazil}

\author{Subhendu Mishra}
\affiliation[Indian Institute of Science]
{Materials Research Centre, Indian Institute of Science, Bangalore 560012, India}

\author{Douglas S Galvao}
\affiliation{Department of Applied Physics and Center for Computational Engineering and Sciences, State
University of Campinas, Campinas, 13083-859, SP, Brazil}

\author{Abhishek Singh}
\affiliation[Indian Institute of Science]
{Materials Research Centre, Indian Institute of Science, Bangalore 560012, India}

\author{Pedro Alves Da Silva Autreto}
\affiliation[Universidade Federal do ABC]
{Center of Natural and Human Sciences, Federal University of ABC, Santo Andre, Sao Paulo, Brazil}

\email{pedro.autreto@ufabc.edu.br}
\phone{+55 (19) 98279-4988}

\title[An \textsf{achemso} demo]
  {Controlling HER activity and stability of $\gamma$- and 6,6,12-Graphyne through engineered B-N doping: DFT and Reactive MD simulations
  }

\abbreviations{Graphyne, DFT, MD, HA}

\begin{document}

\begin{tocentry}
\begin{figure}[H]
    \centering
    \includegraphics[scale=1.0]{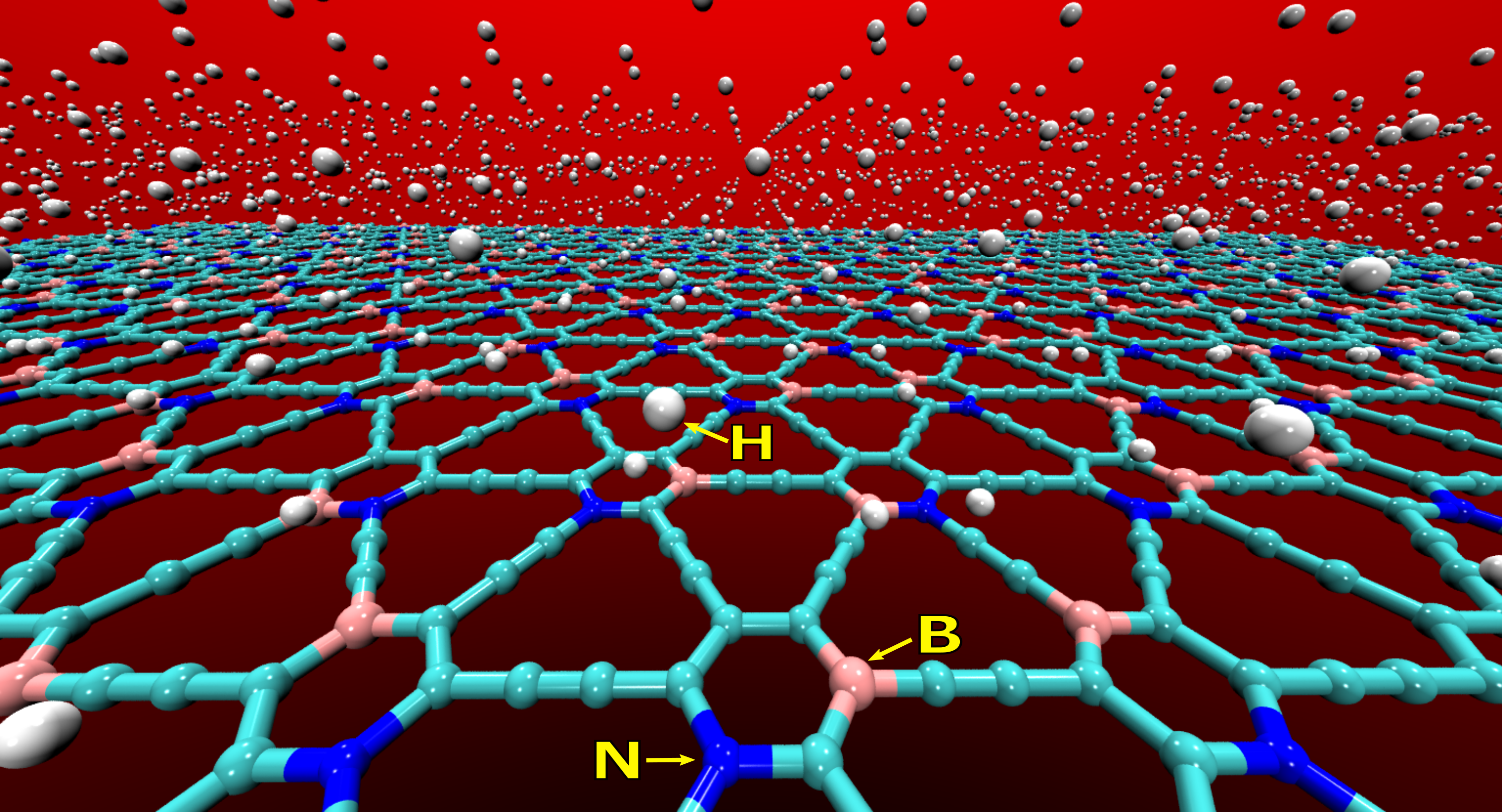}
    \label{fig:toc}
\end{figure}
\end{tocentry}


\begin{abstract}
Graphynes offer a chemically heterogeneous $sp/sp^{2}$ carbon framework with distinct electronic regimes and site-selective reactivity. Here, Density Functional Theory and Reactive Molecular Dynamics Simulations are combined to evaluate pristine, B-doped, N-doped, and B-N co-doped $\gamma$-graphyne and 6,6,12-graphyne (meta/ortho/para). $\gamma$-graphyne is a semiconductor, while 6,6,12-graphyne exhibits an anisotropic Dirac-like semi-metallic dispersion. B/N substitution reconstructs near-$E_F$ states via dopant $\pi$ hybridization, and B-N pairing stabilizes defects through donor-acceptor compensation, with the ortho substitutions being the most favorable. Hydrogen adsorption remains weak on pristine lattices but becomes locally optimized upon doping, with near thermo-neutral $\Delta G_{\mathrm{ads}}$ 'hot spots' predominantly on $sp$-proximate carbon sites adjacent to the dopants. Reactive MD at 300 K further reveals an activity stability trade-off: B-N ortho in $\gamma$-graphyne sustains controlled hydrogen uptake without catastrophic bond scission, whereas B-N meta/para degrade, and 6,6,12-graphyne is generally more susceptible to over-hydrogenation. These results identify the B-N geometry as a key design variable for graphyne-based HER catalysts, which require both a favorable $\Delta G_{\mathrm{ads}}$ and finite-temperature hydrogenation stability.
\end{abstract}

\section{Introduction}
The hydrogen evolution reaction (HER) is a cornerstone process for sustainable H$_2$ production. However, practical applications require catalysts that combine high activity with long-term stability under hydrogen-rich conditions. While noble metals remain reference standards, earth-abundant, metal-free alternatives are actively pursued, and two-dimensional carbon allotropes offer an attractive platform because their electronic structure can be engineered at the atomic scale. Graphynes, a family of $sp/sp^2$ hybridized carbon sheets originally proposed by Baughman and co-workers, are obtained by expanding the graphene topology through the incorporation of acetylenic linkages (C$\equiv$C), creating chemically non-equivalent carbon sites and a tunable $\pi$ network \cite{Baughman1987,Ivanovskii2013a,Kang2019GraphyneAdvances}.\\

Within the graphyne family, $\gamma$-graphyne and 6,6,12-graphyne represent contrasting electronic limits: a semiconducting phase with a finite electronic band gap and an anisotropic Dirac-like semimetal, each based on the topology of the mixed-hybridization lattices \cite{Yun2015First-principles-graphyne,Mirnezhad2012a, desyatkin2022scalable, aliev2025planar}. Such electronically diverse behavior is catalytically relevant because the density and character of states near the Fermi level ($E_F$) regulate charge accommodation and bond formation with adsorbates, thus shaping the attainable adsorption free-energy landscape and its sensitivity to local chemical and mechanical perturbations.\\

For HER, the Sabatier principle favors near thermo-neutral H binding, typically assessed via the adsorption free energy $\Delta G_{\mathrm{ads}}$ \cite{Rossmeisl2005ElectrolysisSurfaces}.
Recent first-principles studies have highlighted graphynes as promising candidates for HER tuning through hetero-atom substitution and related defect engineering, including investigations of 6,6,12-graphyne \cite{Motaghi2025EnhancingStudy,Guo2022}.
In particular, B and N dopants are natural candidates to engineer the low-energy $\pi$ changes: acceptor/donor substitution and dopant–$\pi$ hybridization do not simply shift bands rigidly, but redistribute spectral weight and induce local polarization that can activate adjacent carbon sites more effectively than the hetero-atom itself \cite{Bhattacharya2016TheGraphyne}.
Importantly, such electronic tuning is inseparable from dopant-induced lattice relaxation and symmetry breaking.\\

A recent phonon-based analysis of $\gamma$- and 6,6,12-graphyne demonstrated that B, N, and BN motifs generate inversion-symmetry breaking through bond distortions, leading to lifted mode degeneracies and the appearance of chiral phonons; notably, BN-ortho was identified as a particularly robust configuration from a stability standpoint \cite{Mishra2025ChiralGraphyne}. This provides independent evidence that BN geometry is not a secondary detail but a primary design variable that co-controls charge redistribution and structural stability—two ingredients that must be optimized simultaneously for electrocatalysis. However, two problems remain: (i) how the relative B–N geometry (meta/ortho/para) simultaneously controls defect thermodynamics and the near-$E_F$ electronic texture, and (ii) whether motifs predicted to optimize $\Delta G_{\mathrm{ads}}$ at 0 K remain kinetically robust against hydrogenation-driven bond scission at finite temperature.\\

Here, we address these issues by combining density-functional theory (DFT) with fully atomistic reactive molecular dynamics (ReaxFF) to build a structural stability reactivity map for pristine, B-doped, N-doped, and B-N co-doped $\gamma$-graphyne and 6,6,12-graphyne. We first quantify how isolated and paired dopants affect the electronic band structure and the low-energy states relevant to hydrogen adsorption. We then evaluate defect formation energies and site-resolved $\Delta G_{\mathrm{ads}}$ across chemically distinct $sp$ and $sp^2$ structural motifs. 
Finally, we validate the predicted activity descriptors by explicitly simulating hydrogenation at room temperature, allowing us to identify dopant geometries that balance near thermo-neutral H* binding with resistance to degradation under hydrogen-rich conditions \cite{autreto2014site}.

\section{Computational Methods}

\begin{figure}[H]
    \centering
     \includegraphics[width=0.6\linewidth]{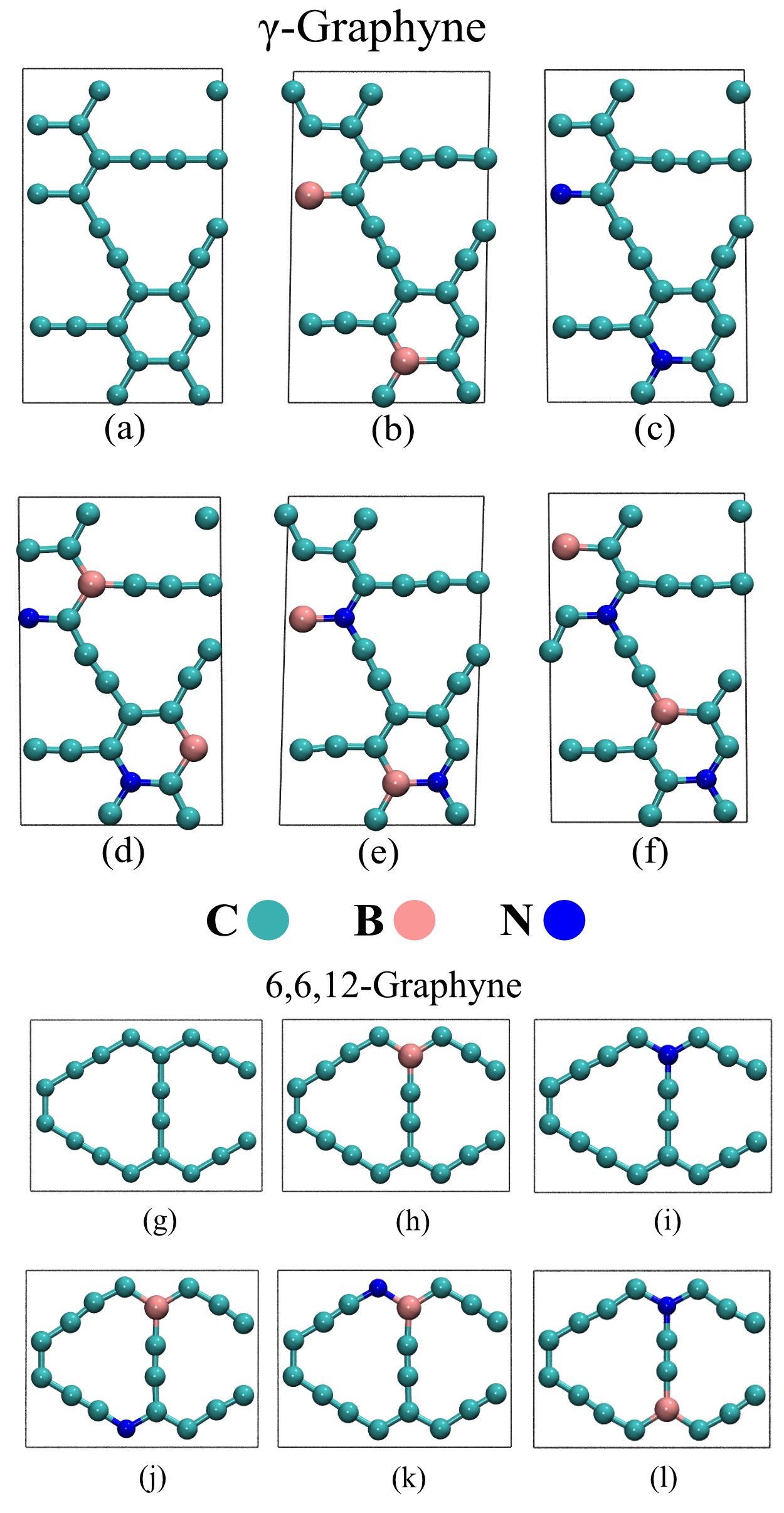}
    \caption{The fully optimized atomic structures of $\gamma$-graphyne (a-f) and 6,6,12-graphyne (g-l) considered in this work. Pristine lattices are shown in (a,g); single substitutional B and N doping at a carbon site in (b,h) and (c,i), respectively; and B-N co-doping in (d,j) meta, (e,k) ortho, and (f,l) para substitutions, defined by the relative placement of the two substitutional dopants.}
    \label{fig:GG_6G_struct}
\end{figure}

To investigate the impact of substitutional B and N doping, as well as B-N co-doping, we constructed the structural models shown in Fig. \ref{fig:GG_6G_struct} for pristine $\gamma$-graphyne and 6,6,12-graphyne and their corresponding doped derivatives. For single doping, a carbon atom within the $C_{6}$ aromatic ring was substitutionally replaced by either B or N, as illustrated in Fig. \ref{fig:GG_6G_struct}(b,c) for $\gamma$-graphyne and Fig. \ref{fig:GG_6G_struct} (h,i) for 6,6,12-graphyne, respectively. For co-doping, a B-N pair was introduced within the same $C_{6}$ ring in three relative arrangements: meta, ortho, and para positions, where the B and N atoms occupy alternating \cite{Mishra2025ChiralGraphyne}, consecutive, and opposite positions in the ring, respectively (Fig. \ref{fig:GG_6G_struct}(d–f) for $\gamma$-graphyne and Fig. \ref{fig:GG_6G_struct}(j-l) for 6,6,12-graphyne).

\subsection{Density Functional Theory Calculations}

Density functional theory (DFT) calculations were performed to assess the effects of B, N, and B-N pair incorporation on the electronic structure and H adsorption properties of $\gamma$-graphyne and 6,6,12-graphyne. All calculations were carried out using the SIESTA code \cite{Soler2002a,Garcia2020}, which employs numerical atomic-orbital basis sets and norm-conserving pseudopotentials to describe the electron-ion interaction \cite{Bylander1982}. The exchange-correlation energy was treated within the generalized gradient approximation using the Perdew–Burke-Ernzerhof (GGA-PBE) functional \cite{Perdew1996}, and a double-zeta plus polarization (DZP) basis set was adopted. Long-range dispersion interactions were included through the DFT-D3 correction \cite{Grimme2010AH-Pu,Garcia2020}. A real-space mesh cutoff of 350 Ry was used, and a vacuum spacing of 20 \AA \ was introduced along the out-of-plane direction to avoid spurious interactions between periodic images.\\

Structural relaxations were performed using a Monkhorst-Pack \cite{Hu2019} 5×5×1 k-point mesh until the maximum residual force on each atom was below 0.02 eV/\AA. Self-consistency was achieved by imposing convergence thresholds of $10^{-4}$ for the density matrix and $10^{-3}$ for the maximum absolute change in the Hamiltonian matrix elements. For electronic band-structure calculations, a Monkhorst–Pack grid dense of 10×10×1 k-points was employed for both $\gamma$-graphyne and 6,6,12-graphyne. Spin-polarized calculations were also performed; however, no appreciable spin imbalance was found, indicating that the optimized structures remain essentially non-magnetic within the present level of theory.\\

To assess the thermodynamic stability of substitutional B and N doping, as well as B–N co-doping, we have computed the dopant formation energy ($E_{\text{form}}$) as: \cite{Liao2022DensityElectrocatalysis,He2019a}:
\begin{equation}
    E_{\text{form}}(\text{ev/atom}) = E_{\text{tot}}(\text{doped}) - E_{\text{tot}}(\text{pristine}) -\sum_i n_{i} \mu_{i},
\end{equation}

where $E_{\text{tot}}(\text{doped})$ and $E_{\text{tot}}(\text{pristine})$ are the total energies of the doped and pristine cells, respectively; $n_{i}$ is the number of atoms of species $i$ added to ($n_{i}$ > 0) or removed from ($n_{i}$ < 0) the pristine lattice to create the doped configuration; and $\mu_i$ is the chemical potential of species $i$, taken from the selected reference reservoirs \cite{Liao2022DensityElectrocatalysis,He2019a}. $\alpha$-B (bulk), the N$_{2}$ molecule, and graphite were used as reservoirs for calculating the formation energy values. Physically, $E_{\text{form}}$ measures the energetic cost (or gain) of creating a defect/dopant configuration relative to the pristine network while exchanging atoms with external reservoirs \cite{Liao2022DensityElectrocatalysis,He2019a}. Therefore, a lower (more negative) formation energy indicates that the doped structure is thermodynamically more favorable under the assumed chemical environment, whereas a higher (positive) $E_{\text{form}}$ implies that incorporation is less favorable and may require non-equilibrium and/or energy-assisted synthesis conditions \cite{Liao2022DensityElectrocatalysis,He2019a}. It should be emphasized that $E_{\text{form}}$ depends on the choice of chemical potentials, i.e., on the experimental growth conditions implicitly represented by the reference states.\\

To quantify the interaction of H atoms with the doped graphyne surfaces, we evaluated the Gibbs free energy of hydrogen adsorption ($\Delta G_{\mathrm{ads}}$) within the computational hydrogen electrode framework \cite{Nrskov2005TrendsEvolution,Nrskov2009TowardsCatalysts}. 
It was calculated as:
\begin{equation}
    \Delta G_{\text{ads}} = \Delta E_{\text{H}}^{*} + \Delta E_{\text{ZPE}} - T \Delta S \approx \Delta E_{\text{H}}^{*} + 0.24 \ \text{eV},
\end{equation}
where the adsorption energy term is:
\begin{equation}
    \Delta E_{\text{H}}^{*} = E_{\text{tot}}(\text{surface} + \text{H}) - E_{\text{tot}}(\text{surface}) - \frac{1}{2} E_{\text{tot}}(\text{H}_{2}).
\end{equation}

Here, $E_{\text{tot}}(\text{surface} + \text{H})$ and $E_{\text{tot}}(\text{surface})$ are the total energy values of the H-adsorbed and clean surfaces, respectively, and $E_{\text{tot}}(\text{H}_{2})$ is the total energy value of an isolated $\text{H}_{2}$ molecule. The constant 0.24 eV accounts for the combined zero-point energy and entropic contribution ($\Delta E_{\text{ZPE}} - T\Delta S$) under standard conditions, a commonly adopted approximation in HER modeling \cite{Nrskov2005TrendsEvolution,Nrskov2009TowardsCatalysts}.

From a physical standpoint view, $\Delta G_{\text{ads}}$ is the key thermodynamic descriptor for HER activity: an optimal catalyst should bind H neither too strongly nor too weakly \cite{Medford2015FromCatalysis,Nrskov2009TowardsCatalysts}. If $\Delta G_{\text{ads}}$ is significantly negative, H binds too strongly and surface sites can become blocked (hindering H$_{2}$ formation/desorption). If $\Delta G_{\text{ads}}$ is significantly positive, H adsorption is unfavorable, and the surface cannot efficiently populate H$^{*}$ intermediates. Consequently, the regime $\Delta G_{\text{ads}} \approx 0$ eV is often considered thermo-neutral and is associated with the best compromise between adsorption and desorption kinetics, consistent with the Sabatier principle and the well-known volcano-type trends for HER catalysts \cite{Liao2022DensityElectrocatalysis,Nrskov2009TowardsCatalysts}.

\subsection{Molecular Dynamics Simulations}

Additionally, reactive molecular dynamics (MD) simulations were carried out using the LAMMPS package \cite{Thompson2022LAMMPSScales} to probe the interaction of atomic hydrogen with pristine and B-, N-, and B-N-doped graphyne sheets, and to assess their finite-temperature structural stability under a hydrogen-rich environment. The dynamics were described using the reactive force field ReaxFF \cite{Aktulga2012ParallelTechniques,VanDuin2001ReaxFF:Hydrocarbons,Fantauzzi2015SurfacePt111}, which captures bond breaking/formation and is therefore appropriate to model hydrogenation and saturation processes in carbon-based frameworks \cite{Marinho2021b}. Charge transfer effects were accounted for through on-the-fly charge equilibration during the MD trajectories (QEq scheme within ReaxFF).\\

The MD simulations were performed in a simulation cell of approximately $100~\text{\AA} \times 100~\text{\AA}$ in-plane, with the graphyne sheet exposed to an atmosphere of atomic H placed in regions above and below the surface. To prevent nonphysical evaporation of the gas phase, hydrogen atoms were confined within a finite volume using reflective boundaries. To mimic the experimental conditions and avoid border effects, a finite, movable graphyne fragment is deposited on a hollow substrate \cite{oliveira2023tetra}.  All trajectories were computed for $500$ ps in the canonical (NVT) ensemble using a Nosé-Hoover thermostat at $T=300$ K \cite{Evans1985TheThermostat}, with a timestep of $0.25$ fs. This finite-temperature, dynamical protocol provides additional validation of the stability of the proposed doped configurations and enables a consistent comparison of their relative hydrogen-uptake/saturation kinetics under identical thermodynamic conditions.

\section{Results and Discussions}

\begin{figure}[H]
    \centering
    \includegraphics[width=1.0\linewidth]{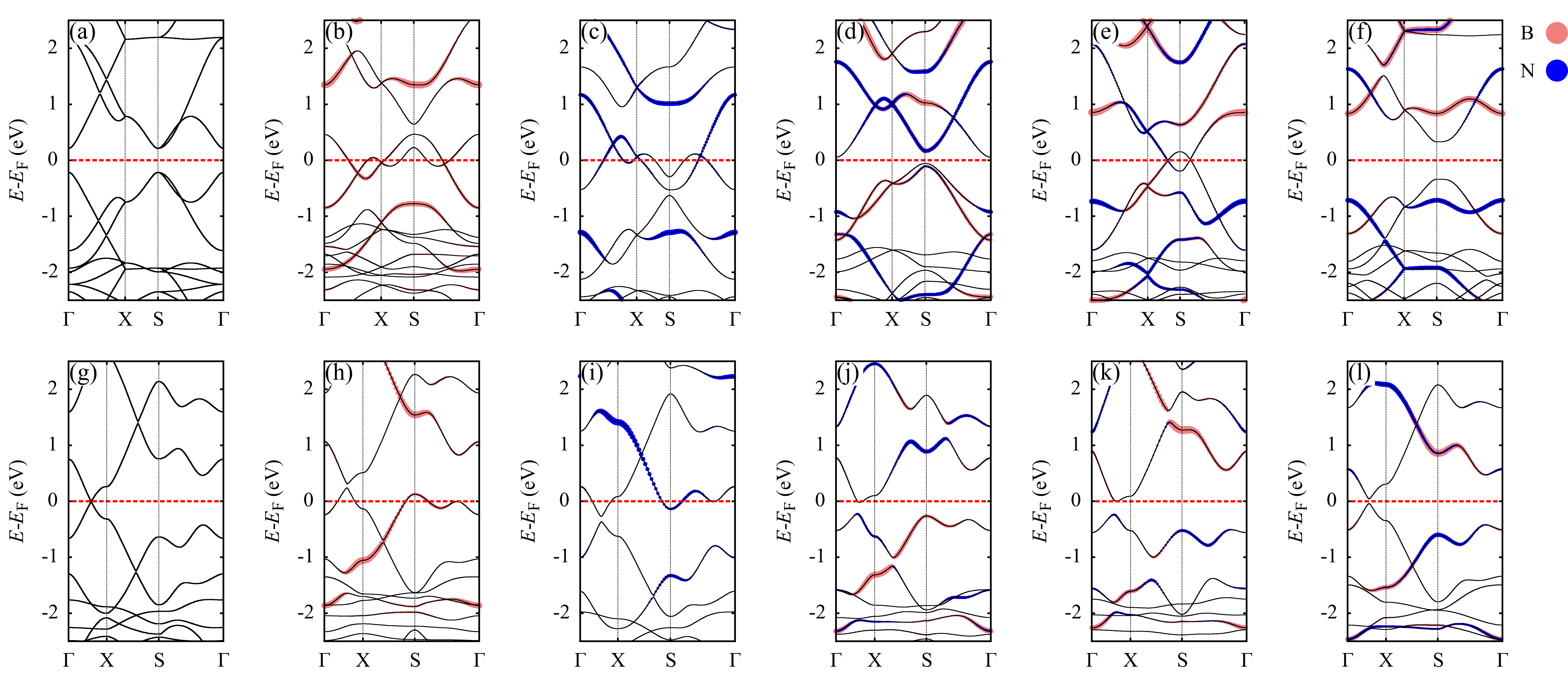}
    \caption{Electronic band structures of $\gamma$-graphyne (a-f) and 6,6,12-graphyne (g-l) for pristine, single-doped, and co-doped configurations, including atomic-orbital projections on the dopants. Pristine lattices are shown in (a,g); substitutional B and N doping (C$\rightarrow$B/N) in (b,h) and (c,i), respectively; and B–N co-doping in meta (d,j), ortho (e,k), and para (f,l) substitutions. The colored markers highlight the projected contributions from B (light-coral) and N (blue) states to the corresponding bands, while black lines denote the total band dispersion. Energies are referenced to the Fermi level ($E_F=0$, red dashed line).}
    \label{fig:GG_6G_bands}
\end{figure}

Figure \ref{fig:GG_6G_bands} summarizes the electronic band structures of pristine and doped $\gamma$-graphyne and 6,6,12-graphyne, including dopant-projected contributions. These two lattices capture two canonical regimes within the graphyne family, semiconductor ($\gamma$-graphyne) and an anisotropic Dirac-like semimetal (6,6,12-graphyne) originating from the mixed $sp/sp^2$ topology that modulates $\pi$-conjugation and enforces direction-dependent electronic dispersions \cite{Baughman1987,Kang2019GraphyneAdvances}. The results for the pristine structures reproduce these limits, with an electronic band gap for $\gamma$-graphyne (Fig. \ref{fig:GG_6G_bands}a) and a Dirac-like crossing near $E_F$ for 6,6,12-graphyne (Fig. \ref{fig:GG_6G_bands}g). These distinct electronic features are catalytically relevant because the density and character of the states around $E_F$ control charge accommodation and bond formation with adsorbates, thereby shaping the accessible $\Delta G_{\mathrm{ads}}$ landscape.

Single substitutional doping perturbs the band edges and near-$E_F$ states in a manner consistent with electron counting and explicit dopant-$\pi$ hybridization. B substitution (less one electron) introduces acceptor-like features and shifts the spectral weight toward $E_F$, with clear B-projected contributions in both cases (Fig. \ref{fig:GG_6G_bands}(b,h)), consistent with prior first-principles analyses of site-dependent B doping in graphynes \cite{Bhattacharya2016TheGraphyne,Kang2016ImportanceGraphynes}. Conversely, N substitution (one electron more) yields donor-like character and generates N-derived dispersive states near the band edges and close to $E_F$ (Fig. \ref{fig:GG_6G_bands}(c,i)), indicating that doping is not a rigid-band effect but a genuine reconstruction of the low-energy manifold \cite{Bhattacharya2016TheGraphyne,Kang2016ImportanceGraphynes}.\\

B-N co-doping further broadens the tuning space by combining partial charge compensation with local symmetry breaking controlled by the B-N arrangement (meta/ortho/para). Consistent with earlier DFT studies on B-N pair doping in graphyne-like networks, the B-N motif can reduce the net carrier imbalance while still introducing a geometry-dependent perturbation that splits bands and reshapes the near-$E_F$ spectrum through a local dipole and anisotropic hybridization \cite{Yun2017DFTNanosheets,Kang2019GraphyneAdvances}. This is particularly evident in 6,6,12-graphyne, where B-N co-doping modifies the Dirac-like dispersion (Fig. \ref{fig:GG_6G_bands}(j-l)), indicating an efficient route to engineer low-energy electronic texture without introducing metallic dopants.\\

In addition to chemical doping, Figs. S1 and S2 show that moderate uniaxial tensile strain (3\% and 5\% along $x$ or $y$) provides a secondary control parameter to fine-tune near-$E_F$ features by re-normalizing bond lengths and effective hoppings. For $\gamma$-graphyne, strain modulates the electronic band gap with a clear direction dependence (Fig.~S1(a,b)), while preserving the dispersive character of dopant-hybridized bands (Fig.~S1(c-l)). For 6,6,12-graphyne, strain distorts the anisotropic Dirac-like dispersion and shifts the crossing region (Fig.~S2(a,b)), with a configuration-dependent response in doped/co-doped cases (Fig.~S2(c-l)). This is consistent with the known strain sensitivity of graphyne-family electronic structures \cite{Yue2013MechanicalPredictions}. Overall, Fig.~2 (and Figs.~S1-S2) establishes doping (B/N/B-N) as the primary effect to engineer near-$E_F$ states, while moderate strain offers complementary fine control, setting the stage for the adsorption thermodynamics discussed next.\\

\begin{figure}[H]
    \centering
    \includegraphics[width=1.0\linewidth]{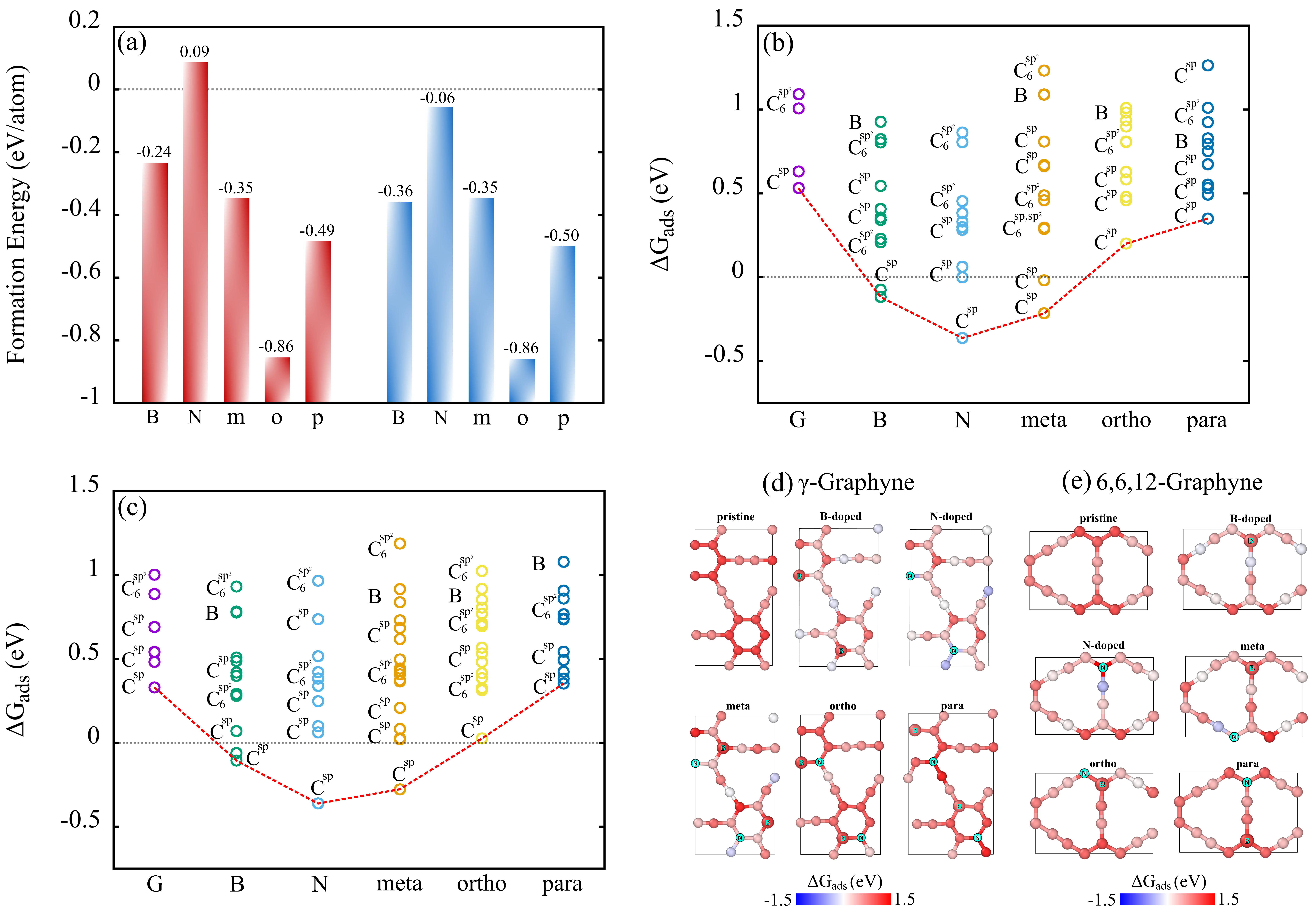} 
    \caption{(a) Formation energy values (in $\mathrm{eV/atom}$) for substitutional B, N, and B-N pair doping in $\gamma$-graphyne (red) and 6,6,12-graphyne (blue). (b,c) Hydrogen adsorption free energies $\Delta G_{\mathrm{ads}}$ (in $\mathrm{eV}$) for pristine (G), B-doped (B), N-doped (N), and B-N co-doped (meta/ortho/para) structures. Each marker denotes a distinct adsorption site, annotated by the local carbon hybridization ($sp$ or $sp^{2}$), and the red dashed line marks the thermo-neutral condition $\Delta G_{\mathrm{ads}}=0$. (d,e) Spatial distribution of the hydrogen adsorption free energy ($\Delta G_{\mathrm{ads}}$) for $\gamma$-graphyne and 6,6,12-graphyne, respectively. The color scale ranges from $-1.5$ eV (blue) to $+1.5$ eV (red).}
    \label{fig:dGads_resumo}
\end{figure}

Figures \ref{fig:dGads_resumo} and Fig. S3 provide a structure, stability, and reactivity link for doped graphyne lattices by combining real-space charge-density fingerprints (Fig.~S3) with defect thermodynamics (Fig. \ref{fig:dGads_resumo}a) and HER descriptors (Fig. \ref{fig:dGads_resumo}(b-e)). 
For the pristine networks, the charge-density isosurfaces highlight the intrinsic electronic inhomogeneity of mixed-$sp/sp^2$ networks: the acetylenic $sp$ chains sustain a more directional, bond-centered charge accumulation than the $sp^2$ rings, consistent with the chemically non-equivalent environments widely reported across graphyne families \cite{Kang2019GraphyneAdvances,Bhattacharya2016TheGraphyne}.\\

Upon B or N substitution (Fig.~S3), the charge density becomes locally polarized and redistributed over neighboring C atoms, evidencing dopant-$\pi$ hybridization rather than a purely electrostatic (rigid-band) perturbation. Importantly, the induced polarization extends into adjacent $sp$-connected motifs, which are known to be particularly responsive to hetero-atom doping in graphynes \cite{Kang2016ImportanceGraphynes,Chen2015WhyStudy}. 
B-N co-doping amplifies this effect in a geometry-dependent manner (meta/ortho/para), consistent with the formation of a B-N induced local dipole and symmetry breaking that reshapes the electronic landscape of nearby carbon sites \cite{Yun2017DFTNanosheets}.

The formation energies in Fig. \ref{fig:dGads_resumo}a confirm that B-N pairing is systematically stabilized relative to isolated B or N defects, with ortho substitution as the most favorable configuration. This trend follows the established donor-acceptor charge-compensation picture for B/N co-doping in graphyne-like lattices, where the B-N arrangement controls both local relaxation and the degree of electronic polarization \cite{Chen2015WhyStudy,Yun2017DFTNanosheets}.\\

Figures \ref{fig:dGads_resumo}(b,c) show that pristine $\gamma$-graphyne and 6,6,12-graphyne remain overall sub-optimal for HER under the commonly adopted thermodynamic target $|\Delta G_{\rm ads}|$  $\leq$ 0.2 eV. Most adsorption sites lie on the weak-binding side ($\Delta G_{\rm ads}>0$~eV), with 6,6,12-graphyne systematically closer to the optimum value than $\gamma$-graphyne. A robust trend across all configurations is that the most active positions are predominantly associated with $sp$-type carbons, whereas $sp^2$ ring sites tend to bind H weakly, consistent with prior graphyne electrocatalysis studies where acetylenic backbones provide chemically “softer” motifs for H$^{*}$ stabilization \cite{Gao2019DopingElectrocatalysis,Guo2022,Motaghi2025EnhancingStudy,Ullah2021HighInsight}.\\

Single B or N substitution markedly reshapes the adsorption landscape in both cases, generating new sites that approach the target window. Importantly, this would improve $\Delta G_{\rm ads}$ not only in the immediate vicinity of the dopant but also at sites beyond the first coordination shell. This non-local response is evident in the dispersion of points in Fig. \ref{fig:dGads_resumo}(b,c) and is further supported by the full adsorption heat maps in Fig. \ref{fig:dGads_resumo}(d,e), which show that the most favorable sites emerge along specific regions of the lattice rather than at a single isolated atom. Such behavior is physically expected in graphynes because dopant-induced charge redistribution and $\pi$-network re-hybridization propagate along the conjugated framework, modulating the reactivity of multiple $sp$-connected motifs. In contrast, adsorption directly at N remains thermodynamically disfavored (large positive $\Delta G_{\rm ads}$) in essentially all cases, indicating that N primarily acts as an electronic perturbation rather than a chemically active H-binding center.\\

The apparent catalytic inertness of the substitutional N site originates from an unfavorable electronic balance for N-H bond formation. Owing to the higher electronegativity of N, its 2p states are stabilized relative to the carbon $\pi$ manifold, which limits effective charge accommodation and orbital matching with the H-1s state. Consistently, the Crystal Orbital Overlap Population (COOP) analyses (Fig. S4) indicate that, unlike the neighboring C-H bond that accumulates bonding contributions predominantly below $E_F$, the metastable N-H geometry shifts a significant fraction of bonding weight to the unoccupied manifold. This reflects an electronically underfilled N-H interaction that does not yield net stabilization upon relaxation, explaining the absence of a true metastable N-H adsorption state. Consequently, N acts primarily as an electronic perturbation that polarizes and re-hybridizes the surrounding conjugated framework, rendering adjacent carbon sites-particularly those connected to the sp backbone the thermodynamically accessible adsorption motifs, with a $\Delta G^{\text{min}}_{\rm ads}\sim -$0.36 eV for both $\gamma$-graphyne and 6,6,12-graphyne.\\

B-N co-doping introduces an additional geometric degree of freedom that controls whether the system moves toward the Sabatier window or into over/under-binding regimes. For $\gamma$-graphyne (Fig. \ref{fig:dGads_resumo}b), the meta substitution yields the strongest stabilization (including slightly negative $\Delta G_{\rm ads}$ values, i.e., a tendency to overbind), whereas the ortho substitution provides the closest approach to the target range among the B-N geometries. The para substitution remains weak-binding, with all sites shifted to positive $\Delta G^{min}_{\rm ads} \sim 0.35$~eV, outside the optimal window. For 6,6,12-graphyne, B-N co-doping can place selected $sp$-proximate sites directly inside the thermo-neutral window (e.g., $\Delta G^{min}_{\rm ads} \sim 0.03$~eV for ortho substitution), while the para configuration remains substantially underactive ($\Delta G^{min}_{\rm ads} \sim 0.35$~eV). The spatial patterns in Fig. \ref{fig:dGads_resumo}(d,e) corroborate that these near-optimal sites cluster around B-N perturbed regions but can also extend along the $sp$ backbone. This, in turn, reinforces the idea that the catalytically relevant motif is a BN-polarized carbon environment rather than the heteroatoms themselves.\\

Finally, these real-space adsorption trends connect directly to Fig. \ref{fig:GG_6G_bands}: the dopant-projected, near-$E_F$ reconstructed manifold provides the electronic reservoir that governs charge accommodation into the H-$1s$ state, thereby setting whether specific $sp$-carbon motifs fall inside ($|\Delta G_{\rm ads}|\leq 0.2$~eV) or outside the Sabatier window.\\

\begin{figure}[H]
    \centering
    \includegraphics[width=1.0\linewidth]{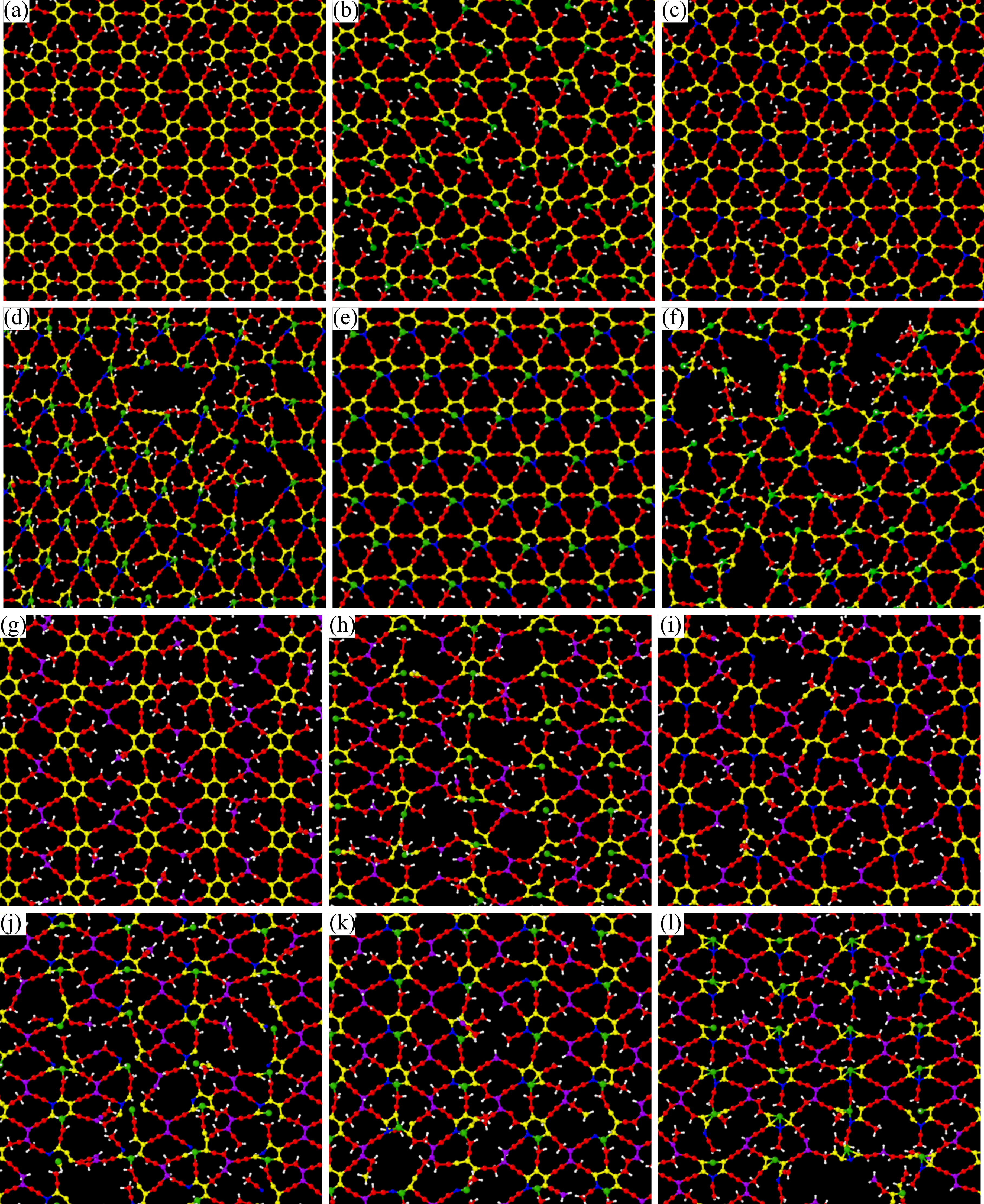}
    \caption{Final atomic configurations obtained after 500 ps of thermalization for $\gamma$-graphyne (a–f) and 6,6,12-graphyne (g–l). Panels (a–f) correspond to $\gamma$-graphyne in the following configurations: pristine (a), B-doped (b), N-doped (c), and BN co-doped in meta (d), ortho (e), and para (f) substitutions. Panels (g–l) show the corresponding final structures for 6,6,12-graphyne in the same sequence: pristine (g), B-doped (h), N-doped (i), and BN co-doped in meta (j), ortho (k), and para (l) substitutions. All structures correspond to the equilibrated atomic positions after molecular dynamics thermalization.}
    \label{fig:MD_last}
\end{figure}

Figure \ref{fig:MD_last} probes the finite-temperature kinetic robustness of pristine and doped graphynes under a hydrogen-rich environment, providing a necessary complement to the 0~K, low-coverage descriptor $\Delta G_{\rm ads}$ (Fig. \ref{fig:dGads_resumo}). 
In $sp$-rich carbon frameworks such as graphynes, sequential H uptake can progressively convert local bonding motifs (e.g., by weakening/re-hybridizing acetylenic linkages). As a result, configurations that appear attractive from an isolated-adsorbate thermodynamic viewpoint may still fail by over-hydrogenation-driven bond scission at finite temperature \cite{Peng2014,Tan2012,Kang2019GraphyneAdvances}. 
Therefore, the MD snapshots in Fig. \ref{fig:MD_last} act as a kinetic "filter" for practical catalyst design, where structural integrity under realistic hydrogen exposure is a strict requirement in addition to near thermo-neutral adsorption.\\

For $\gamma$-graphyne (Fig.\ref{fig:MD_last}(a-f)), the pristine lattice and single B/N substitutions remain structurally intact after 500~ps at 300~K, indicating that moderate electronic activation of the $sp/sp^2$ backbone can be accommodated without significant structural deformations. 
In contrast, B-N co-doping exhibits a sharp geometry dependence: the meta and para substitutions display clear bond cleavage. This is consistent with the fact that B-N arrangements can generate strongly localized polarization patterns (Fig.~S3) that promote spatially concentrated H uptake (Fig.~S4), thereby accelerating over-hydrogenation of vulnerable acetylenic motifs.\\

Remarkably, B-N ortho substitution preserves the lattice integrity, suggesting that this specific B-N placement yields a more 'distributed' perturbation (electronic and structural), capable of sustaining adsorption events without triggering re-hybridization and scission. This result is consistent with the broader understanding that graphyne stability and reactivity are governed by the interplay between acetylenic-link activity and the ability of the network to dissipate local perturbations \cite{Kang2016ImportanceGraphynes,Mirnezhad2012a}.\\

The stability landscape is substantially more restrictive for 6,6,12-graphyne (Fig. \ref{fig:MD_last}(g-l)). Even the pristine sheet shows incipient bond-breaking, and single B/N substitutions increase the extent of breakings, while B-N co-doping leads to pronounced network deformations. This trend is consistent with the intrinsically higher chemical responsiveness of anisotropic, $sp$-rich graphyne backbones and the enhanced sensitivity of rectangular-lattice graphynes to local symmetry breaking and charge redistribution \cite{Kang2019GraphyneAdvances,Bhattacharya2016TheGraphyne}. Collectively, Fig. \ref{fig:MD_last} reveals a clear network- and geometry-dependent activity and stability trade-off: motifs that strongly activate the electronic landscape near the B-N defect can also transform the lattice beyond a hydrogenation stability threshold.\\

\begin{figure}[H]
    \centering
    \includegraphics[width=1.0\linewidth]{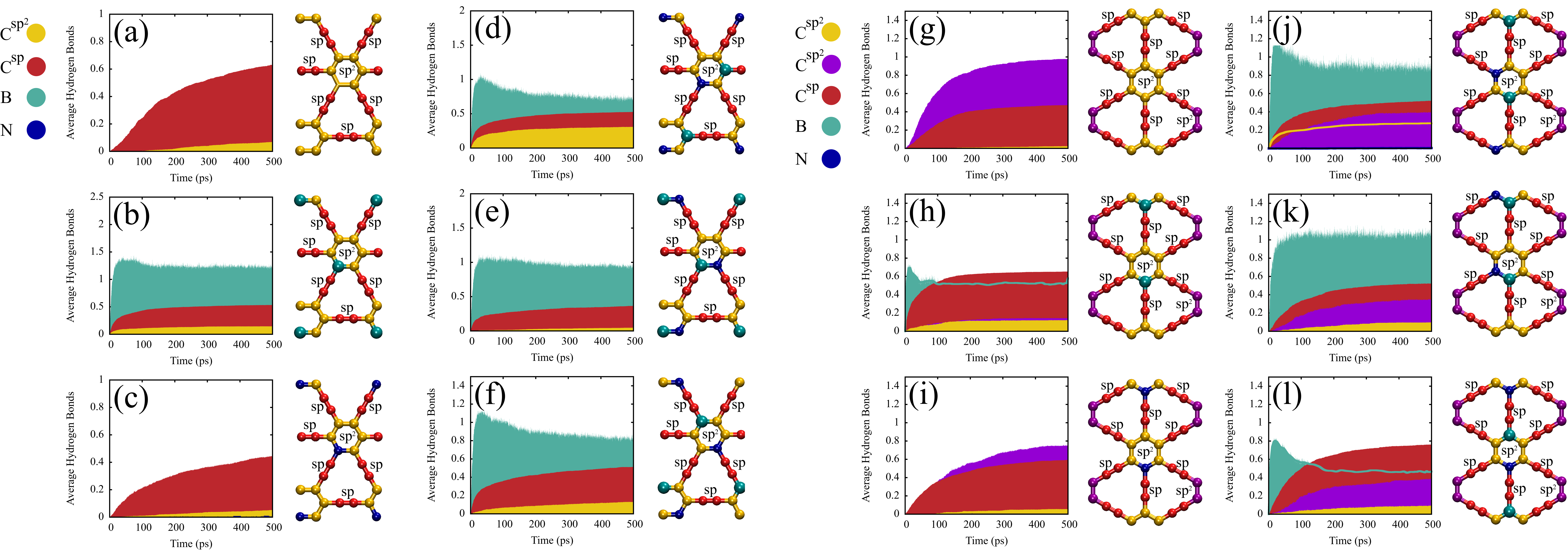}
    \caption{Time evolution of the average number of hydrogen bonds obtained from molecular-dynamics simulations for $\gamma$-graphyne (a-f) and 6,6,12-graphyne (g-l). For each system, panels correspond to the same set of structures considered in Fig. 1: pristine, single substitutional B and N doping, and B-N co-doping in meta/ortho/para substitutions (ordered as a-f and g-l, respectively). The stacked areas report the site-resolved contributions to the hydrogen-bond statistics from the different atomic species and local environments (C in $sp$ and $sp^{2}$ coordination, and dopant B/N; color code as indicated). The structural snapshots next to each plot identify the adsorption site(s) and the corresponding carbon hybridization.}
    \label{fig:MD_GG_6G}
\end{figure}

Figure \ref{fig:MD_GG_6G} quantifies the corresponding hydrogenation kinetics by tracking the time evolution of H-bond formation and resolving the contributions from $sp$ vs $sp^2$ carbon environments and dopant-centered motifs. For $\gamma$-graphyne (Fig. \ref{fig:MD_GG_6G}(a-f)), pristine hydrogen uptake is dominated by $sp$ carbons, in line with the higher intrinsic activity of acetylenic units reported for graphyne-based electrocatalyst platforms \cite{Gao2019DopingElectrocatalysis,Guo2022}. B-substitution accelerates early-time uptake and reaches a relatively well-defined plateau, consistent with B acting as an efficient H-trapping/activation motif while still maintaining structural integrity (Fig. \ref{fig:MD_last}b). N-substitution produces a more modest uptake, indicating weaker kinetic driving toward high coverage under identical conditions. Most importantly, B-N co-doping again displays a geometry-controlled crossover: meta and para substitutions show large and evolving H-bond counts without a clearly stabilized plateau, consistent with progressive over-hydrogenation that correlates directly with the bond scission observed in Fig. \ref{fig:MD_last}(d,f). By contrast, B-N ortho configuration combines sustained uptake with a comparatively stable plateau-like behavior, supporting a controlled saturation regime rather than structural collapse, in agreement with its preserved final structure (Fig. \ref{fig:MD_last}e).\\

For 6,6,12-graphyne (Fig. \ref{fig:MD_GG_6G}(g-l)), H-bond accumulation is faster and larger even in the pristine case, and dopants—particularly B-N pairs drive a rapid rise to high, persistent H-bond counts. This kinetic signature is characteristic of over-hydrogenation conditions and is consistent with the severe structural degradation observed in Fig. \ref{fig:MD_last}(j-l). Taken together, Figs. \ref{fig:MD_last}-\ref{fig:MD_GG_6G} provide a mechanistic bridge from the static thermodynamics (Fig. \ref{fig:dGads_resumo}) to dynamical robustness: dopant motifs that create highly active H-binding environments can either enable controlled saturation (e.g., $\gamma$-graphyne B-N ortho) or trigger uncontrolled hydrogenation leading to bond scission (notably in B-N co-doped 6,6,12-graphyne). This combined evidence strengthens the central design criterion that practical graphyne-based HER candidates must satisfy both near thermo-neutral $\Delta G_{\rm ads}$ and kinetic stability against hydrogenation-driven degradation.

\section{Conclusions}

We have established a structural, stability and reactivity picture for B, N, and B-N (meta/ortho/para) doping in $\gamma$-graphyne and 6,6,12-graphyne by combining electronic band structures, defect formation energies, $\Delta G_{\mathrm{ads}}$ descriptors, and reactive MD at 300 K. The pristine lattices span two limiting electronic regimes (gapped $\gamma$-graphyne vs anisotropic Dirac-like 6,6,12-graphyne). Dopant-projected bands show that B/N actively hybridize with the $\pi$ network and reshape the near-$E_F$ states relevant for adsorption. B-N co-doping is systematically stabilized relative to isolated dopants, consistent with donor-acceptor compensation, with B-N ortho being the most favorable geometry. Doping converts weak H-binding into a strongly site-selective landscape where near thermo-neutral adsorption emerges at specific carbon motifs adjacent to the dopants, particularly along the $sp$ backbone. This indicates that B-N primarily activates neighboring C sites rather than acting as a hetero-atom-centered active site. Reactive MD reveals a decisive kinetic filter: hydrogenation can induce over-coverage and bond scission in a network- and geometry-dependent manner. $\gamma$-graphyne remains robust for pristine/single-doped cases and uniquely sustains controlled saturation for B-N ortho, whereas B-N meta/para degrade; 6,6,12-graphyne is intrinsically more prone to bond breaking, which becomes severe under B-N co-doping. Overall, B-N geometry emerges as a key design variable for graphyne-based HER catalysts, which must jointly satisfy near thermo-neutral $\Delta G_{\mathrm{ads}}$ and kinetic robustness under hydrogen-rich conditions. While near thermo-neutral $\Delta G_{\mathrm{ads}}$ is a necessary thermodynamic criterion, the overall HER performance is also governed by the kinetic barriers along the Volmer/Heyrovsky/Tafel pathways, which will be addressed in future work.

\bibliography{arxiv/references2}

\providecommand{\latin}[1]{#1}
\makeatletter
\providecommand{\doi}
  {\begingroup\let\do\@makeother\dospecials
  \catcode`\{=1 \catcode`\}=2 \doi@aux}
\providecommand{\doi@aux}[1]{\endgroup\texttt{#1}}
\makeatother
\providecommand*\mcitethebibliography{\thebibliography}
\csname @ifundefined\endcsname{endmcitethebibliography}  {\let\endmcitethebibliography\endthebibliography}{}
\begin{mcitethebibliography}{40}
\providecommand*\natexlab[1]{#1}
\providecommand*\mciteSetBstSublistMode[1]{}
\providecommand*\mciteSetBstMaxWidthForm[2]{}
\providecommand*\mciteBstWouldAddEndPuncttrue
  {\def\EndOfBibitem{\unskip.}}
\providecommand*\mciteBstWouldAddEndPunctfalse
  {\let\EndOfBibitem\relax}
\providecommand*\mciteSetBstMidEndSepPunct[3]{}
\providecommand*\mciteSetBstSublistLabelBeginEnd[3]{}
\providecommand*\EndOfBibitem{}
\mciteSetBstSublistMode{f}
\mciteSetBstMaxWidthForm{subitem}{(\alph{mcitesubitemcount})}
\mciteSetBstSublistLabelBeginEnd
  {\mcitemaxwidthsubitemform\space}
  {\relax}
  {\relax}

\bibitem[Baughman \latin{et~al.}(1987)Baughman, Eckhardt, and Kertesz]{Baughman1987}
Baughman,~R.~H.; Eckhardt,~H.; Kertesz,~M. {Structure‐property predictions for new planar forms of carbon: Layered phases containing sp2 and sp atoms}. \emph{The Journal of Chemical Physics} \textbf{1987}, \emph{87}, 6687--6699\relax
\mciteBstWouldAddEndPuncttrue
\mciteSetBstMidEndSepPunct{\mcitedefaultmidpunct}
{\mcitedefaultendpunct}{\mcitedefaultseppunct}\relax
\EndOfBibitem
\bibitem[Ivanovskii(2013)]{Ivanovskii2013a}
Ivanovskii,~A.~L. {Graphynes and graphdyines}. \emph{Progress in Solid State Chemistry} \textbf{2013}, \emph{41}, 1--19\relax
\mciteBstWouldAddEndPuncttrue
\mciteSetBstMidEndSepPunct{\mcitedefaultmidpunct}
{\mcitedefaultendpunct}{\mcitedefaultseppunct}\relax
\EndOfBibitem
\bibitem[Kang \latin{et~al.}(2019)Kang, Wei, and Li]{Kang2019GraphyneAdvances}
Kang,~J.; Wei,~Z.; Li,~J. {Graphyne and Its Family: Recent Theoretical Advances}. \emph{ACS Applied Materials and Interfaces} \textbf{2019}, \emph{11}, 2692--2706\relax
\mciteBstWouldAddEndPuncttrue
\mciteSetBstMidEndSepPunct{\mcitedefaultmidpunct}
{\mcitedefaultendpunct}{\mcitedefaultseppunct}\relax
\EndOfBibitem
\bibitem[Yun \latin{et~al.}(2015)Yun, Zhang, Yan, and Zhao]{Yun2015First-principles-graphyne}
Yun,~J.; Zhang,~Z.; Yan,~J.; Zhao,~W. {First-principles study of B doping effect on the electronic structure and magnetic properties of {$\gamma$}-graphyne}. \emph{Thin Solid Films} \textbf{2015}, \emph{589}, 662--668\relax
\mciteBstWouldAddEndPuncttrue
\mciteSetBstMidEndSepPunct{\mcitedefaultmidpunct}
{\mcitedefaultendpunct}{\mcitedefaultseppunct}\relax
\EndOfBibitem
\bibitem[Mirnezhad \latin{et~al.}(2012)Mirnezhad, Ansari, Rouhi, Seifi, and Faghihnasiri]{Mirnezhad2012a}
Mirnezhad,~M.; Ansari,~R.; Rouhi,~H.; Seifi,~M.; Faghihnasiri,~M. {Mechanical properties of two-dimensional graphyne sheet under hydrogen adsorption}. \emph{Solid State Communications} \textbf{2012}, \emph{152}, 1885--1889\relax
\mciteBstWouldAddEndPuncttrue
\mciteSetBstMidEndSepPunct{\mcitedefaultmidpunct}
{\mcitedefaultendpunct}{\mcitedefaultseppunct}\relax
\EndOfBibitem
\bibitem[Desyatkin \latin{et~al.}(2022)Desyatkin, Martin, Aliev, Chapman, Fonseca, Galv{\~a}o, Miller, Stone, Wang, Zakhidov, \latin{et~al.} others]{desyatkin2022scalable}
Desyatkin,~V.~G.; Martin,~W.~B.; Aliev,~A.~E.; Chapman,~N.~E.; Fonseca,~A.~F.; Galv{\~a}o,~D.~S.; Miller,~E.~R.; Stone,~K.~H.; Wang,~Z.; Zakhidov,~D.; others Scalable synthesis and characterization of multilayer $\gamma$-graphyne, new carbon crystals with a small direct band gap. \emph{Journal of the American Chemical Society} \textbf{2022}, \emph{144}, 17999--18008\relax
\mciteBstWouldAddEndPuncttrue
\mciteSetBstMidEndSepPunct{\mcitedefaultmidpunct}
{\mcitedefaultendpunct}{\mcitedefaultseppunct}\relax
\EndOfBibitem
\bibitem[Aliev \latin{et~al.}(2025)Aliev, Guo, Fonseca, Razal, Wang, Galv{\~a}o, Bolding, Chapman-Wilson, Desyatkin, Leisen, \latin{et~al.} others]{aliev2025planar}
Aliev,~A.~E.; Guo,~Y.; Fonseca,~A.~F.; Razal,~J.~M.; Wang,~Z.; Galv{\~a}o,~D.~S.; Bolding,~C.~M.; Chapman-Wilson,~N.~E.; Desyatkin,~V.~G.; Leisen,~J.~E.; others A planar-sheet nongraphitic zero-bandgap sp2 carbon phase made by the low-temperature reaction of $\gamma$-graphyne. \emph{Proceedings of the National Academy of Sciences} \textbf{2025}, \emph{122}, e2413194122\relax
\mciteBstWouldAddEndPuncttrue
\mciteSetBstMidEndSepPunct{\mcitedefaultmidpunct}
{\mcitedefaultendpunct}{\mcitedefaultseppunct}\relax
\EndOfBibitem
\bibitem[Rossmeisl \latin{et~al.}(2005)Rossmeisl, Logadottir, and N{\o}rskov]{Rossmeisl2005ElectrolysisSurfaces}
Rossmeisl,~J.; Logadottir,~A.; N{\o}rskov,~J.~K. {Electrolysis of water on (oxidized) metal surfaces}. \emph{Chemical Physics} \textbf{2005}, \emph{319}, 178--184\relax
\mciteBstWouldAddEndPuncttrue
\mciteSetBstMidEndSepPunct{\mcitedefaultmidpunct}
{\mcitedefaultendpunct}{\mcitedefaultseppunct}\relax
\EndOfBibitem
\bibitem[Motaghi and Mohammadi-Manesh(2025)Motaghi, and Mohammadi-Manesh]{Motaghi2025EnhancingStudy}
Motaghi,~F.; Mohammadi-Manesh,~H. {Enhancing hydrogen evolution reaction using transition metal atoms on 6,6,12-graphyne: A DFT study}. \emph{Journal of the Serbian Chemical Society} \textbf{2025}, \emph{90}, 609--621\relax
\mciteBstWouldAddEndPuncttrue
\mciteSetBstMidEndSepPunct{\mcitedefaultmidpunct}
{\mcitedefaultendpunct}{\mcitedefaultseppunct}\relax
\EndOfBibitem
\bibitem[Guo \latin{et~al.}(2022)Guo, Ji, and Cui]{Guo2022}
Guo,~M.; Ji,~M.; Cui,~W. {Theoretical investigation of HER/OER/ORR catalytic activity of single atom-decorated graphyne by DFT and comparative DOS analyses}. \emph{Applied Surface Science} \textbf{2022}, \emph{592}, 153237\relax
\mciteBstWouldAddEndPuncttrue
\mciteSetBstMidEndSepPunct{\mcitedefaultmidpunct}
{\mcitedefaultendpunct}{\mcitedefaultseppunct}\relax
\EndOfBibitem
\bibitem[Bhattacharya and Sarkar(2016)Bhattacharya, and Sarkar]{Bhattacharya2016TheGraphyne}
Bhattacharya,~B.; Sarkar,~U. {The effect of boron and nitrogen doping in electronic, magnetic, and optical properties of graphyne}. \emph{Journal of Physical Chemistry C} \textbf{2016}, \emph{120}, 26793--26806\relax
\mciteBstWouldAddEndPuncttrue
\mciteSetBstMidEndSepPunct{\mcitedefaultmidpunct}
{\mcitedefaultendpunct}{\mcitedefaultseppunct}\relax
\EndOfBibitem
\bibitem[Mishra \latin{et~al.}(2025)Mishra, Chakraborty, Galvao, Autreto, and Singh]{Mishra2025ChiralGraphyne}
Mishra,~S.; Chakraborty,~A.; Galvao,~D.~S.; Autreto,~P. A.~S.; Singh,~A.~K. {Chiral Phonons in Graphyne}. \emph{The Journal of Physical Chemistry C} \textbf{2025}, \relax
\mciteBstWouldAddEndPunctfalse
\mciteSetBstMidEndSepPunct{\mcitedefaultmidpunct}
{}{\mcitedefaultseppunct}\relax
\EndOfBibitem
\bibitem[Autreto \latin{et~al.}(2014)Autreto, De~Sousa, and Galvao]{autreto2014site}
Autreto,~P.; De~Sousa,~J.; Galvao,~D. Site-dependent hydrogenation on graphdiyne. \emph{Carbon} \textbf{2014}, \emph{77}, 829--834\relax
\mciteBstWouldAddEndPuncttrue
\mciteSetBstMidEndSepPunct{\mcitedefaultmidpunct}
{\mcitedefaultendpunct}{\mcitedefaultseppunct}\relax
\EndOfBibitem
\bibitem[Soler \latin{et~al.}(2002)Soler, Artacho, Gale, Garc{\'{i}}a, Junquera, Ordej{\'{o}}n, and S{\'{a}}nchez-Portal]{Soler2002a}
Soler,~J.~M.; Artacho,~E.; Gale,~J.~D.; Garc{\'{i}}a,~A.; Junquera,~J.; Ordej{\'{o}}n,~P.; S{\'{a}}nchez-Portal,~D. {The SIESTA method for ab initio order- N materials simulation}. \emph{Journal of Physics: Condensed Matter} \textbf{2002}, \emph{14}, 2745--2779\relax
\mciteBstWouldAddEndPuncttrue
\mciteSetBstMidEndSepPunct{\mcitedefaultmidpunct}
{\mcitedefaultendpunct}{\mcitedefaultseppunct}\relax
\EndOfBibitem
\bibitem[Garc{\'{i}}a \latin{et~al.}(2020)Garc{\'{i}}a, Papior, Akhtar, Artacho, Blum, Bosoni, Brandimarte, Brandbyge, Cerd{\'{a}}, Corsetti, Cuadrado, Dikan, Ferrer, Gale, Garc{\'{i}}a-Fern{\'{a}}ndez, Garc{\'{i}}a-Su{\'{a}}rez, Garc{\'{i}}a, Huhs, Illera, Koryt{\'{a}}r, Koval, Lebedeva, Lin, L{\'{o}}pez-Tarifa, Mayo, Mohr, Ordej{\'{o}}n, Postnikov, Pouillon, Pruneda, Robles, S{\'{a}}nchez-Portal, Soler, Ullah, Yu, and Junquera]{Garcia2020}
Garc{\'{i}}a,~A. \latin{et~al.}  {Siesta: Recent developments and applications}. \emph{The Journal of Chemical Physics} \textbf{2020}, \emph{152}\relax
\mciteBstWouldAddEndPuncttrue
\mciteSetBstMidEndSepPunct{\mcitedefaultmidpunct}
{\mcitedefaultendpunct}{\mcitedefaultseppunct}\relax
\EndOfBibitem
\bibitem[Bylander and Kleinman(1982)Bylander, and Kleinman]{Bylander1982}
Bylander,~D.~M.; Kleinman,~L. {Efficacious Form for Model Pseudopotentials}. \emph{Physical Review Letters} \textbf{1982}, \emph{48}, 1425--1428\relax
\mciteBstWouldAddEndPuncttrue
\mciteSetBstMidEndSepPunct{\mcitedefaultmidpunct}
{\mcitedefaultendpunct}{\mcitedefaultseppunct}\relax
\EndOfBibitem
\bibitem[Perdew \latin{et~al.}(1996)Perdew, Burke, and Ernzerhof]{Perdew1996}
Perdew,~J.~P.; Burke,~K.; Ernzerhof,~M. {Generalized gradient approximation made simple}. \emph{Physical Review Letters} \textbf{1996}, \emph{77}, 3865--3868\relax
\mciteBstWouldAddEndPuncttrue
\mciteSetBstMidEndSepPunct{\mcitedefaultmidpunct}
{\mcitedefaultendpunct}{\mcitedefaultseppunct}\relax
\EndOfBibitem
\bibitem[Grimme \latin{et~al.}(2010)Grimme, Antony, Ehrlich, and Krieg]{Grimme2010AH-Pu}
Grimme,~S.; Antony,~J.; Ehrlich,~S.; Krieg,~H. {A consistent and accurate ab initio parametrization of density functional dispersion correction (DFT-D) for the 94 elements H-Pu}. \emph{Journal of Chemical Physics} \textbf{2010}, \emph{132}\relax
\mciteBstWouldAddEndPuncttrue
\mciteSetBstMidEndSepPunct{\mcitedefaultmidpunct}
{\mcitedefaultendpunct}{\mcitedefaultseppunct}\relax
\EndOfBibitem
\bibitem[Monkhorst and Pack(1976)Monkhorst, and Pack]{Hu2019}
Monkhorst,~H.~J.; Pack,~J.~D. {Special points for Brillouin-zone integrations}. \emph{Physical Review B} \textbf{1976}, \emph{13}, 5188--5192\relax
\mciteBstWouldAddEndPuncttrue
\mciteSetBstMidEndSepPunct{\mcitedefaultmidpunct}
{\mcitedefaultendpunct}{\mcitedefaultseppunct}\relax
\EndOfBibitem
\bibitem[Liao \latin{et~al.}(2022)Liao, Lu, Xia, Liu, Wang, Zhao, Wang, and Zhao]{Liao2022DensityElectrocatalysis}
Liao,~X.; Lu,~R.; Xia,~L.; Liu,~Q.; Wang,~H.; Zhao,~K.; Wang,~Z.; Zhao,~Y. {Density Functional Theory for Electrocatalysis}. \emph{Energy and Environmental Materials} \textbf{2022}, \emph{5}, 157--185\relax
\mciteBstWouldAddEndPuncttrue
\mciteSetBstMidEndSepPunct{\mcitedefaultmidpunct}
{\mcitedefaultendpunct}{\mcitedefaultseppunct}\relax
\EndOfBibitem
\bibitem[He \latin{et~al.}(2019)He, Yu, Li, and Zhao]{He2019a}
He,~Q.; Yu,~B.; Li,~Z.; Zhao,~Y. {Density Functional Theory for Battery Materials}. \emph{Energy {\&} Environmental Materials} \textbf{2019}, \emph{2}, 264--279\relax
\mciteBstWouldAddEndPuncttrue
\mciteSetBstMidEndSepPunct{\mcitedefaultmidpunct}
{\mcitedefaultendpunct}{\mcitedefaultseppunct}\relax
\EndOfBibitem
\bibitem[N{\o}rskov \latin{et~al.}(2005)N{\o}rskov, Bligaard, Logadottir, Kitchin, Chen, Pandelov, and Stimming]{Nrskov2005TrendsEvolution}
N{\o}rskov,~J.~K.; Bligaard,~T.; Logadottir,~A.; Kitchin,~J.~R.; Chen,~J.~G.; Pandelov,~S.; Stimming,~U. {Trends in the Exchange Current for Hydrogen Evolution}. \emph{Journal of The Electrochemical Society} \textbf{2005}, \emph{152}, J23\relax
\mciteBstWouldAddEndPuncttrue
\mciteSetBstMidEndSepPunct{\mcitedefaultmidpunct}
{\mcitedefaultendpunct}{\mcitedefaultseppunct}\relax
\EndOfBibitem
\bibitem[N{\o}rskov \latin{et~al.}(2009)N{\o}rskov, Bligaard, Rossmeisl, and Christensen]{Nrskov2009TowardsCatalysts}
N{\o}rskov,~J.~K.; Bligaard,~T.; Rossmeisl,~J.; Christensen,~C.~H. {Towards the computational design of solid catalysts}. \emph{Nature Chemistry} \textbf{2009}, \emph{1}, 37--46\relax
\mciteBstWouldAddEndPuncttrue
\mciteSetBstMidEndSepPunct{\mcitedefaultmidpunct}
{\mcitedefaultendpunct}{\mcitedefaultseppunct}\relax
\EndOfBibitem
\bibitem[Medford \latin{et~al.}(2015)Medford, Vojvodic, Hummelsh{\o}j, Voss, Abild-Pedersen, Studt, Bligaard, Nilsson, and N{\o}rskov]{Medford2015FromCatalysis}
Medford,~A.~J.; Vojvodic,~A.; Hummelsh{\o}j,~J.~S.; Voss,~J.; Abild-Pedersen,~F.; Studt,~F.; Bligaard,~T.; Nilsson,~A.; N{\o}rskov,~J.~K. {From the Sabatier principle to a predictive theory of transition-metal heterogeneous catalysis}. \emph{Journal of Catalysis} \textbf{2015}, \emph{328}, 36--42\relax
\mciteBstWouldAddEndPuncttrue
\mciteSetBstMidEndSepPunct{\mcitedefaultmidpunct}
{\mcitedefaultendpunct}{\mcitedefaultseppunct}\relax
\EndOfBibitem
\bibitem[Thompson \latin{et~al.}(2022)Thompson, Aktulga, Berger, Bolintineanu, Brown, Crozier, in~'t Veld, Kohlmeyer, Moore, Nguyen, Shan, Stevens, Tranchida, Trott, and Plimpton]{Thompson2022LAMMPSScales}
Thompson,~A.~P.; Aktulga,~H.~M.; Berger,~R.; Bolintineanu,~D.~S.; Brown,~W.~M.; Crozier,~P.~S.; in~'t Veld,~P.~J.; Kohlmeyer,~A.; Moore,~S.~G.; Nguyen,~T.~D.; Shan,~R.; Stevens,~M.~J.; Tranchida,~J.; Trott,~C.; Plimpton,~S.~J. {LAMMPS - a flexible simulation tool for particle-based materials modeling at the atomic, meso, and continuum scales}. \emph{Computer Physics Communications} \textbf{2022}, \emph{271}, 108171\relax
\mciteBstWouldAddEndPuncttrue
\mciteSetBstMidEndSepPunct{\mcitedefaultmidpunct}
{\mcitedefaultendpunct}{\mcitedefaultseppunct}\relax
\EndOfBibitem
\bibitem[Aktulga \latin{et~al.}(2012)Aktulga, Fogarty, Pandit, and Grama]{Aktulga2012ParallelTechniques}
Aktulga,~H.~M.; Fogarty,~J.~C.; Pandit,~S.~A.; Grama,~A.~Y. {Parallel reactive molecular dynamics: Numerical methods and algorithmic techniques}. \emph{Parallel Computing} \textbf{2012}, \emph{38}, 245--259\relax
\mciteBstWouldAddEndPuncttrue
\mciteSetBstMidEndSepPunct{\mcitedefaultmidpunct}
{\mcitedefaultendpunct}{\mcitedefaultseppunct}\relax
\EndOfBibitem
\bibitem[Van~Duin \latin{et~al.}(2001)Van~Duin, Dasgupta, Lorant, and Goddard]{VanDuin2001ReaxFF:Hydrocarbons}
Van~Duin,~A.~C.; Dasgupta,~S.; Lorant,~F.; Goddard,~W.~A. {ReaxFF: A reactive force field for hydrocarbons}. \emph{Journal of Physical Chemistry A} \textbf{2001}, \emph{105}, 9396--9409\relax
\mciteBstWouldAddEndPuncttrue
\mciteSetBstMidEndSepPunct{\mcitedefaultmidpunct}
{\mcitedefaultendpunct}{\mcitedefaultseppunct}\relax
\EndOfBibitem
\bibitem[Fantauzzi \latin{et~al.}(2015)Fantauzzi, Mueller, Sabo, van Duin, and Jacob]{Fantauzzi2015SurfacePt111}
Fantauzzi,~D.; Mueller,~J.~E.; Sabo,~L.; van Duin,~A. C.~T.; Jacob,~T. {Surface Buckling and Subsurface Oxygen: Atomistic Insights into the Surface Oxidation of Pt(111)}. \emph{ChemPhysChem} \textbf{2015}, \emph{16}, 2797--2802\relax
\mciteBstWouldAddEndPuncttrue
\mciteSetBstMidEndSepPunct{\mcitedefaultmidpunct}
{\mcitedefaultendpunct}{\mcitedefaultseppunct}\relax
\EndOfBibitem
\bibitem[Marinho and da~Silva~Autreto(2021)Marinho, and da~Silva~Autreto]{Marinho2021b}
Marinho,~E.; da~Silva~Autreto,~P.~A. {Me-graphane: tailoring the structural and electronic properties of Me-grapheneviahydrogenation}. \emph{Physical Chemistry Chemical Physics} \textbf{2021}, \emph{23}, 9483--9491\relax
\mciteBstWouldAddEndPuncttrue
\mciteSetBstMidEndSepPunct{\mcitedefaultmidpunct}
{\mcitedefaultendpunct}{\mcitedefaultseppunct}\relax
\EndOfBibitem
\bibitem[Oliveira \latin{et~al.}(2023)Oliveira, Medina, Galvao, and Autreto]{oliveira2023tetra}
Oliveira,~C.~C.; Medina,~M.; Galvao,~D.~S.; Autreto,~P.~A. Tetra-penta-deca-hexagonal-graphene (TPDH-graphene) hydrogenation patterns: dynamics and electronic structure. \emph{Physical Chemistry Chemical Physics} \textbf{2023}, \emph{25}, 13088--13093\relax
\mciteBstWouldAddEndPuncttrue
\mciteSetBstMidEndSepPunct{\mcitedefaultmidpunct}
{\mcitedefaultendpunct}{\mcitedefaultseppunct}\relax
\EndOfBibitem
\bibitem[Evans and Holian(1985)Evans, and Holian]{Evans1985TheThermostat}
Evans,~D.~J.; Holian,~B.~L. {The Nose-Hoover thermostat}. \emph{The Journal of Chemical Physics} \textbf{1985}, \emph{83}, 4069--4074\relax
\mciteBstWouldAddEndPuncttrue
\mciteSetBstMidEndSepPunct{\mcitedefaultmidpunct}
{\mcitedefaultendpunct}{\mcitedefaultseppunct}\relax
\EndOfBibitem
\bibitem[Kang \latin{et~al.}(2016)Kang, Shi, Wang, and Lee]{Kang2016ImportanceGraphynes}
Kang,~B.; Shi,~H.; Wang,~F.~F.; Lee,~J.~Y. {Importance of doping site of B, N, and O in tuning electronic structure of graphynes}. \emph{Carbon} \textbf{2016}, \emph{105}, 156--162\relax
\mciteBstWouldAddEndPuncttrue
\mciteSetBstMidEndSepPunct{\mcitedefaultmidpunct}
{\mcitedefaultendpunct}{\mcitedefaultseppunct}\relax
\EndOfBibitem
\bibitem[Yun \latin{et~al.}(2017)Yun, Zhang, Xu, Yan, Zhao, and Zhang]{Yun2017DFTNanosheets}
Yun,~J.; Zhang,~Y.; Xu,~M.; Yan,~J.; Zhao,~W.; Zhang,~Z. {DFT study of the effect of BN pair doping on the electronic and optical properties of graphyne nanosheets}. \emph{Journal of Materials Science} \textbf{2017}, \emph{52}, 10294--10307\relax
\mciteBstWouldAddEndPuncttrue
\mciteSetBstMidEndSepPunct{\mcitedefaultmidpunct}
{\mcitedefaultendpunct}{\mcitedefaultseppunct}\relax
\EndOfBibitem
\bibitem[Yue \latin{et~al.}(2013)Yue, Chang, Kang, Qin, and Li]{Yue2013MechanicalPredictions}
Yue,~Q.; Chang,~S.; Kang,~J.; Qin,~S.; Li,~J. {Mechanical and electronic properties of graphyne and its family under elastic strain: Theoretical predictions}. \emph{Journal of Physical Chemistry C} \textbf{2013}, \emph{117}, 14804--14811\relax
\mciteBstWouldAddEndPuncttrue
\mciteSetBstMidEndSepPunct{\mcitedefaultmidpunct}
{\mcitedefaultendpunct}{\mcitedefaultseppunct}\relax
\EndOfBibitem
\bibitem[Chen \latin{et~al.}(2015)Chen, Qiao, An, and Xia]{Chen2015WhyStudy}
Chen,~X.; Qiao,~Q.; An,~L.; Xia,~D. {Why Do Boron and Nitrogen Doped {$\alpha$}- And {$\gamma$}-graphyne exhibit different oxygen reduction mechanism? a first-principles study}. \emph{Journal of Physical Chemistry C} \textbf{2015}, \emph{119}, 11493--11498\relax
\mciteBstWouldAddEndPuncttrue
\mciteSetBstMidEndSepPunct{\mcitedefaultmidpunct}
{\mcitedefaultendpunct}{\mcitedefaultseppunct}\relax
\EndOfBibitem
\bibitem[Gao \latin{et~al.}(2019)Gao, Zhou, Cheng, Tan, Liu, and Shen]{Gao2019DopingElectrocatalysis}
Gao,~X.; Zhou,~Y.; Cheng,~Z.; Tan,~Y.; Liu,~S.; Shen,~Z. {Doping sp-hybridized B atoms in graphyne supported single cobalt atoms for hydrogen evolution electrocatalysis}. \emph{International Journal of Hydrogen Energy} \textbf{2019}, \emph{44}, 27421--27428\relax
\mciteBstWouldAddEndPuncttrue
\mciteSetBstMidEndSepPunct{\mcitedefaultmidpunct}
{\mcitedefaultendpunct}{\mcitedefaultseppunct}\relax
\EndOfBibitem
\bibitem[Ullah \latin{et~al.}(2021)Ullah, Ayub, and Mahmood]{Ullah2021HighInsight}
Ullah,~F.; Ayub,~K.; Mahmood,~T. {High performance SACs for HER process using late first-row transition metals anchored on graphyne support: A DFT insight}. \emph{International Journal of Hydrogen Energy} \textbf{2021}, \emph{46}, 37814--37823\relax
\mciteBstWouldAddEndPuncttrue
\mciteSetBstMidEndSepPunct{\mcitedefaultmidpunct}
{\mcitedefaultendpunct}{\mcitedefaultseppunct}\relax
\EndOfBibitem
\bibitem[Peng \latin{et~al.}(2014)Peng, Dearden, Crean, Han, Liu, Wen, and De]{Peng2014}
Peng,~Q.; Dearden,~A.~K.; Crean,~J.; Han,~L.; Liu,~S.; Wen,~X.; De,~S. {New materials graphyne, graphdiyne, graphone, and graphane: Review of properties, synthesis, and application in nanotechnology}. \emph{Nanotechnology, Science and Applications} \textbf{2014}, \emph{7}, 1--29\relax
\mciteBstWouldAddEndPuncttrue
\mciteSetBstMidEndSepPunct{\mcitedefaultmidpunct}
{\mcitedefaultendpunct}{\mcitedefaultseppunct}\relax
\EndOfBibitem
\bibitem[Tan \latin{et~al.}(2012)Tan, He, and Zhao]{Tan2012}
Tan,~J.; He,~X.; Zhao,~M. {First-principles study of hydrogenated graphyne and its family: Stable configurations and electronic structures}. \emph{Diamond and Related Materials} \textbf{2012}, \emph{29}, 42--47\relax
\mciteBstWouldAddEndPuncttrue
\mciteSetBstMidEndSepPunct{\mcitedefaultmidpunct}
{\mcitedefaultendpunct}{\mcitedefaultseppunct}\relax
\EndOfBibitem
\end{mcitethebibliography}

\end{document}


\begin{suppinfo}

\begin{figure}[H]
    \centering
    \includegraphics[width=1.0\linewidth]{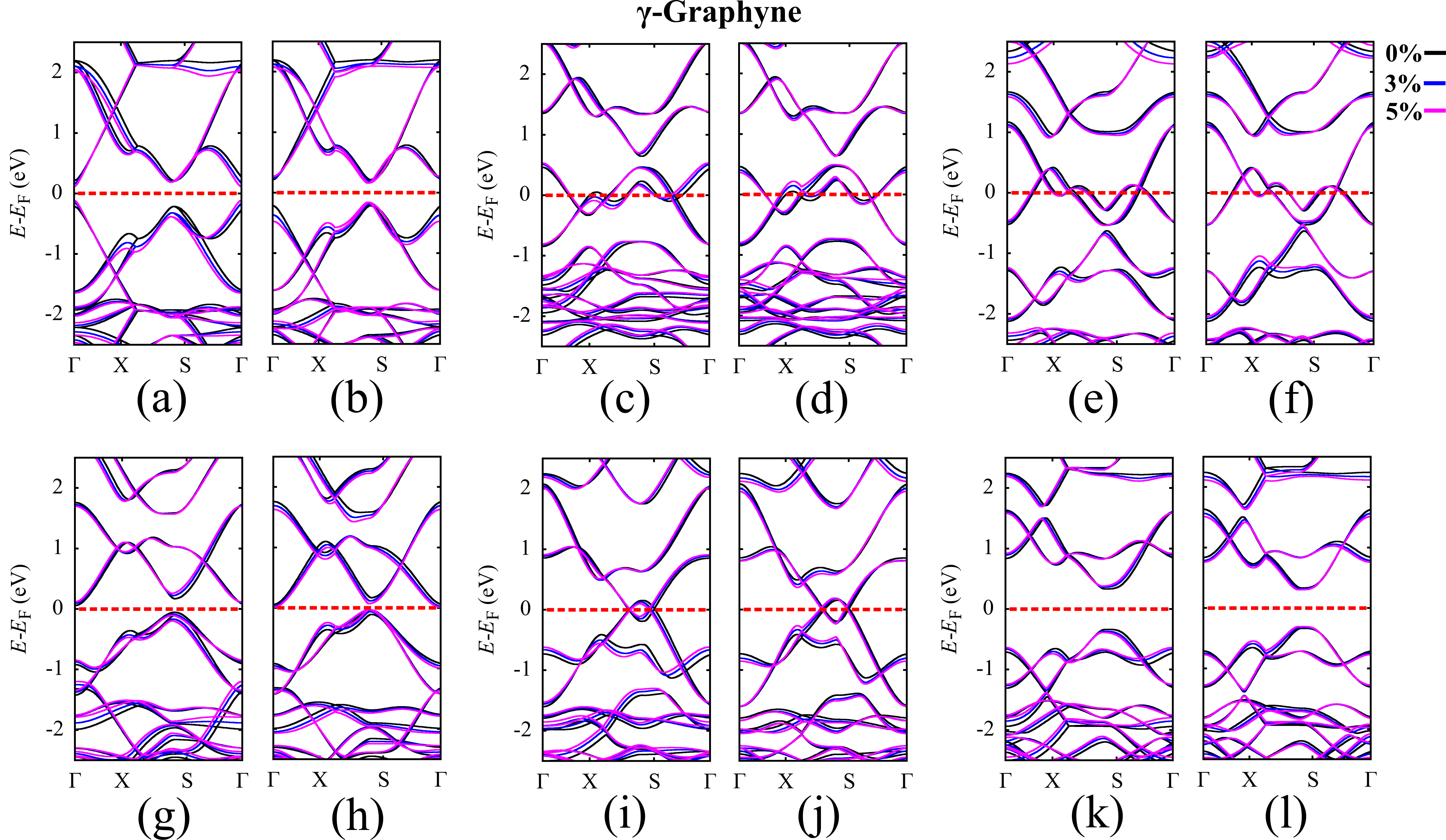}
    \caption*{Figure S1: Electronic band structures of $\gamma$-graphyne under uniaxial strain. Panels (a,b) show the pristine system under strain applied along the $x$ and $y$ directions, respectively; (c,d) B-doped; (e,f) N-doped; (g,h) B--N co-doped in the meta substitution; (i,j) B--N co-doped in the ortho substitution; and (k,l) B--N co-doped in the para substitution. For each case, the band dispersion is shown at 0\% (black), 3\% (blue), and 5\% (magenta) strain. Energies are referenced to the Fermi level ($E_F = 0$, red dashed line).}
    \label{fig:G_G_bands_str}
\end{figure}

\begin{figure}[H]
    \centering
    \includegraphics[width=1.0\linewidth]{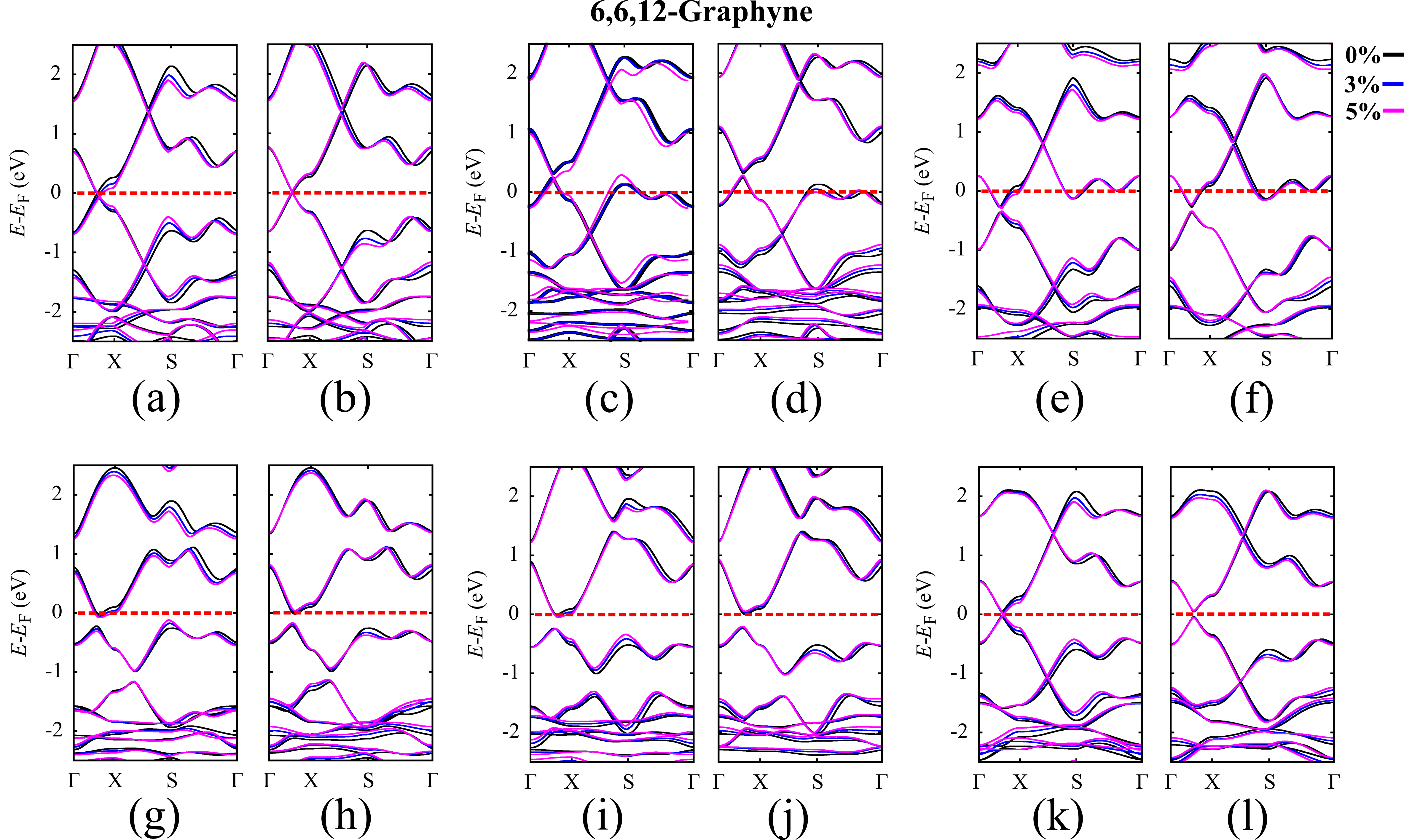}
        \caption*{Figure S2: Electronic band structures of 6,6,12-graphyne under uniaxial strain. Panels (a,b) show the pristine system under strain applied along the $x$ and $y$ directions, respectively; (c,d) B-doped; (e,f) N-doped; (g,h) B--N co-doped in the meta substitution; (i,j) B--N co-doped in the ortho substitution; and (k,l) B--N co-doped in the para substitution. For each case, the band dispersion is shown at 0\% (black), 3\% (blue), and 5\% (magenta) strain. Energies are referenced to the Fermi level ($E_F = 0$, red dashed line).}
    \label{fig:6_G_bands_str}
\end{figure}

\begin{figure}[H]
    \centering
    \includegraphics[width=0.8\linewidth]{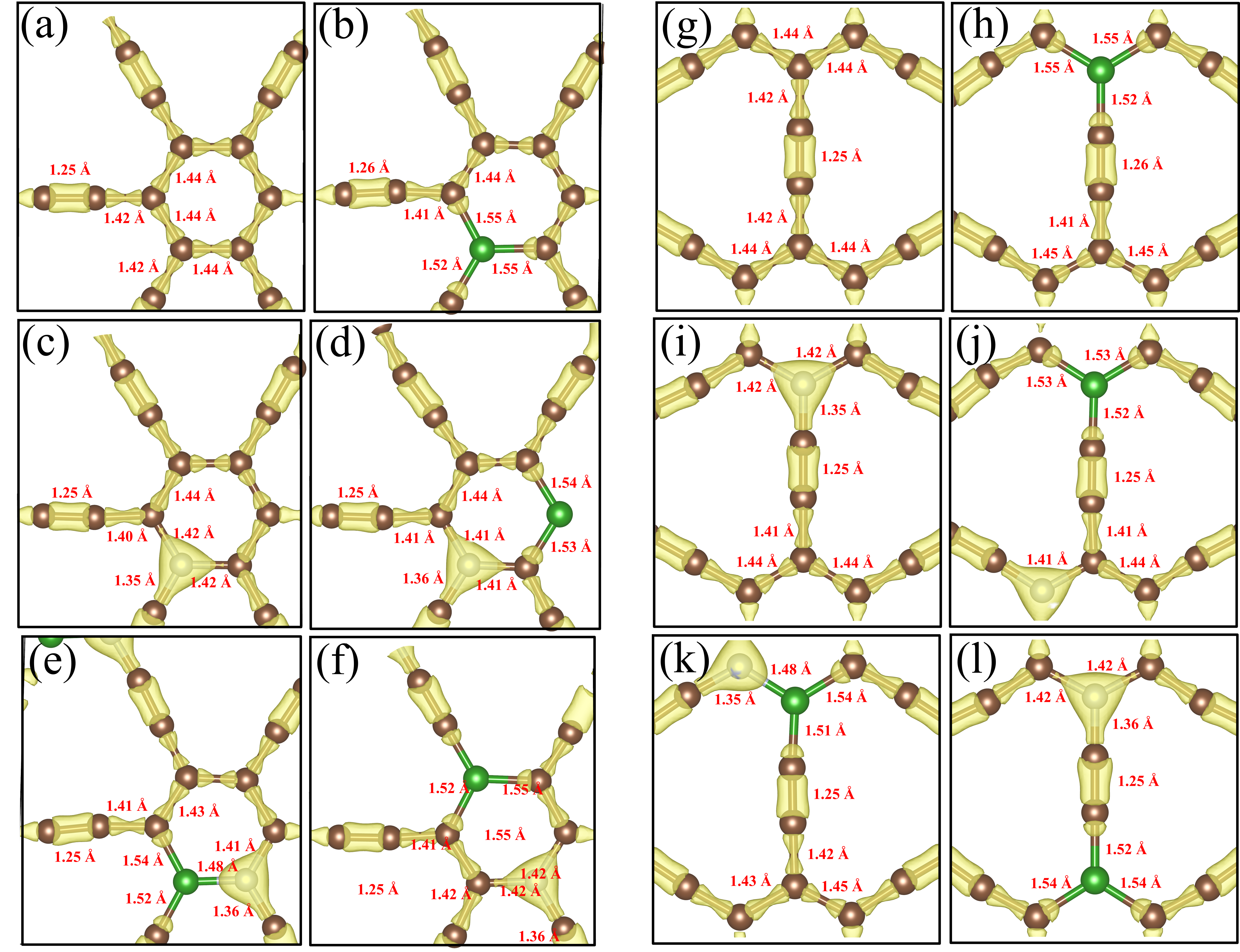}
        \caption*{Figure S3: Charge-density isosurfaces (isovalue: 0.265 e/\AA$^{-3}$) for $\gamma$-graphyne (a--f) and 6,6,12-graphyne (g--l) in the configurations considered in this work. For each allotrope, the panels are shown in the same order adopted in the main text: pristine, B-doped, N-doped, and B--N co-doped in the meta, ortho, and para substitutions. Selected optimized bond lengths (in \AA) are indicated in red to highlight local structural distortions induced by substitutional doping and B--N pair geometry.}
    \label{fig:6_G_bands_str}
\end{figure}


\begin{figure}
    \centering
    \includegraphics[width=0.85\linewidth]{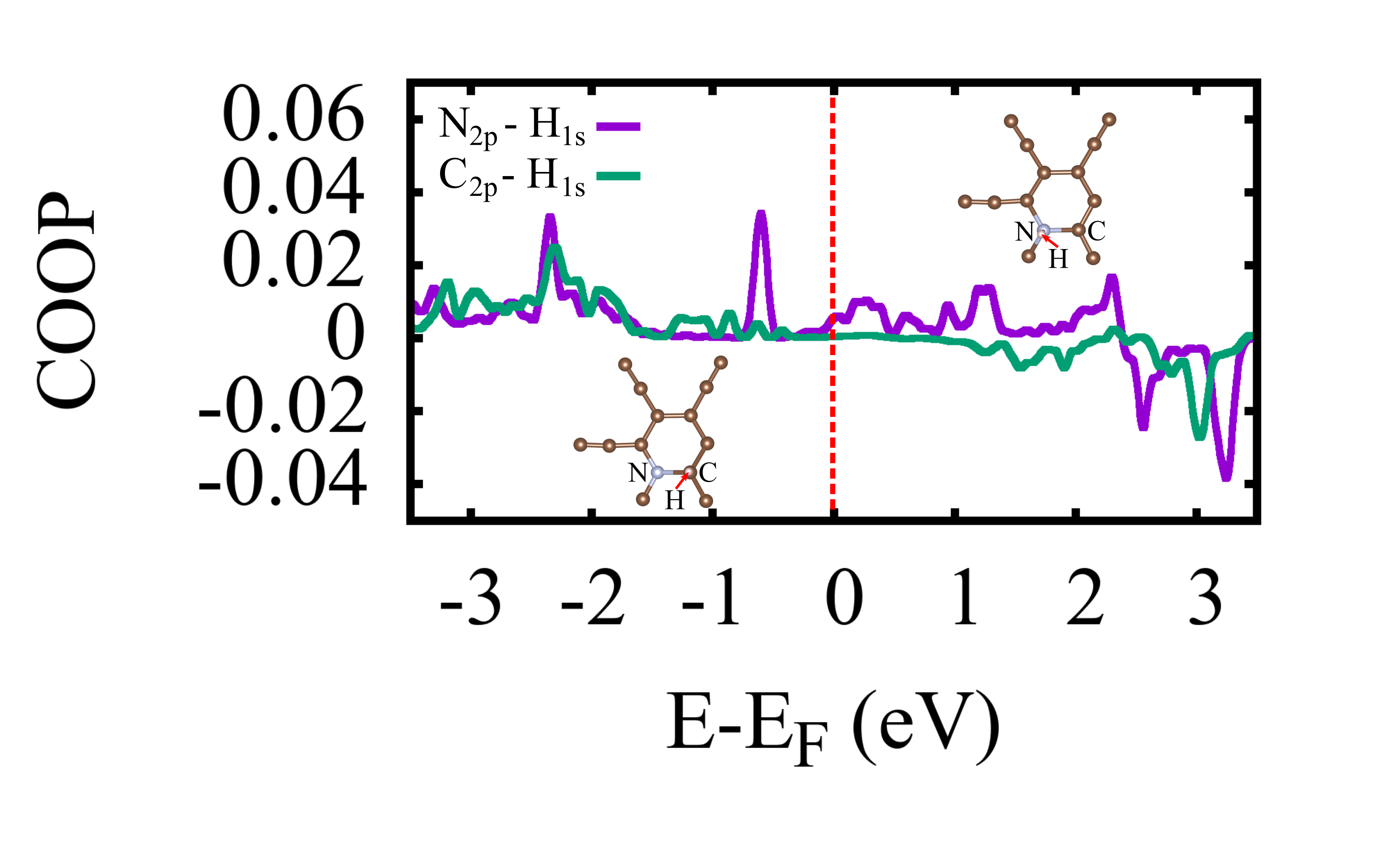}
    \caption*{Figure S4: Crystal Orbital Overlap Population (COOP) projected onto the N${2p}$–H${1s}$ interaction (purple) and the C${2p}$–H${1s}$ interaction (green), where C is the first-neighbor carbon atom adjacent to the substitutional N dopant. Energies are referenced to the Fermi level ($E_F=0$ eV, red dashed line) and reported as $E-E_F$. Positive (negative) COOP values indicate bonding (anti-bonding) contributions. The N–H curve corresponds to a constrained metastable adsorption geometry (no stable N–H minimum was found upon relaxation), whereas the C–H curve corresponds to the preferred adsorption on the neighboring carbon site, consistent with the thermodynamic trend that hydrogen binds to the dopant-polarized carbon environment rather than to the N atom itself. Insets depict the representative adsorption geometries used for each projection.}
    \label{fig:COOP_NC}
\end{figure}



\end{suppinfo}